\documentclass[a4paper,aps,prd,10pt,preprintnumbers,showpacs,twocolumn,superscriptaddress,nofootinbib,amsmath,amssymb,floatfix]{revtex4-1}\def\JCAPstyle#1{}

\usepackage{graphicx}
\usepackage[utf8]{inputenc}
\usepackage[T1]{fontenc}
\usepackage{cmap}

\DeclareMathAlphabet{\pazocal}{OMS}{zplm}{m}{n}

\begin{document}

\preprint{APS/-QNM}

\title{The influence of Aharonov-Casher effect on the generalized Dirac oscillator in the cosmic string space-time}
\author{Hao Chen}
\email{gs.ch19@gzu.edu.cn}
\affiliation{College of Physics, Guizhou University, Guiyang, 550025, China}
\author{Zheng-Wen Long}
\email{ zwlong@gzu.edu.cn}
\affiliation{College of Physics, Guizhou University, Guiyang, 550025, China}
\author{Chao-Yun Long}
\email{ long.chaoyun66@163.com}
\affiliation{College of Physics, Guizhou University, Guiyang, 550025, China}
\author{Soroush Zare}
\email{soroushzrg@gmail.com}
\affiliation{Faculty of Physics, Shahrood University of Technology, Shahrood, Iran}
\author{Hassan Hassanabadi}
\email{h.hasanabadi@shahroodut.ac.ir}
\affiliation{Faculty of Physics, Shahrood University of Technology, Shahrood, Iran}
\date{\today }

\begin{abstract}
In this manuscript, we investigate the influence of the Aharonov-Casher effect on the generalized Dirac oscillator containing the Coulomb-type potential function related to a relativistic neutral particle having a permanent magnetic dipole moment interacting with the external electromagnetic fields in $(1+2)$-dimensional cosmic string space-time. The eigenfunctions and energy eigenvalues of such a Dirac oscillator are derived by using the Nikifornov-Uvarov method.
 We indicate that implementing the scenario gives the relativistic modified exact analytical solutions. In this way, we can see that the degeneracy of the relevant relativistic energy eigenvalues is broken by depending on the Coulomb strength parameter under the influence of the curvature effect and the Aharonov-Casher effect.
\end{abstract}

\maketitle

\section{Introduction}
In the last decades, harmonic interactions have played a crucial role in quantum dynamics to study the behavior of the moving particle subject to the electromagnetic field and scalar potential. Hence, implementing harmonic oscillators in quantum field theory in flat and curved space-time has been a prevalent scenario. Therefore, to conceive and describe harmonic oscillator in the relativistic regime,  Moshinsky and Szczepaniak \cite{re1,re2} presented the well-known coupling model between a relativistic spin-half particle and the harmonic quantum oscillator in the Dirac equation. This model came up according to modifying the momentum operator by considering a nonminimal coupling. The relevant relativistic oscillator was called a Dirac oscillator \cite{CarvalhoEPJC2016,AndradeEPJC2014,re3,re17.2,re4}.\\

We know that the physical system associated with the Dirac oscillator can explained as the electric field interact with anomalous magnetic moment\cite{re4}.
In 2013, a very important experiment for the Dirac oscillator model was confirmed by Franco-Villafa\~{n}e et al. \cite{re5}. Subsequently, the relativistic fermions with spin one-half is widely employed especially in topological defect space-time. Now, let us mention some of the work done in this area, for instance, examining the influence of the Dirac oscillator in presence of the spinning cosmic string space-time \cite{re6,re7}, the fermions described by the Dirac equation interacting with the scalar and vector potentials in the topological defect background \cite{re8,re9,re10}, the Dirac oscillator in the most simple topological defect described by the cosmic string space-time under the background of the rainbow gravity \cite{re11,re12}, the non-inertial effect on the Dirac field with topological defects background \cite{re13,re14}, and the Dirac oscillator under Lorentz symmetry breaking \cite{rer1}. Meanwhile, the Dirac oscillator, generated by the $\kappa$-Poincar\'{e}-algebra, in the context of quantum deformation, is studied in Refs. \cite{AndradePLB2014-1,AndradePLB2014-2}.

Our work is motivated in examining a modified Dirac oscillator interaction related to a neutral particle interacting with the external electromagnetic field under the influences of the Aharonov-Casher (A-C) and curvature effects in $(1+2)$-dimensional cosmic string space-time. By the way, the external electromagnetic fields can generate the Aharonov-Bohm (A-B) \cite{re15} and A-C \cite{re16} effects. In particular,  the effects are often non-negligible in dealing with the Dirac equation problem interacting with an external electromagnetic field in topological defects background. Accordingly, some interesting examples related to the A-B effect had been studied in connection with the fermions interacting with the localized magnetic field in topological defects \cite{re7,re17,re17.1}, and the influence of the A-C effect on Dirac theory in the different scenarios of curved space-time \cite{re18,re19}. Meanwhile, some significant works had also been done based on the extended Dirac oscillator defined by replacing a potential function $f(r)$ instead of the radial coordinate $r$ in cosmic string \cite{re20} and cosmic dislocation space-time \cite{chen1}. Other similar works include the Duffin-Kemmer-Petiau \cite{re21,re23} and Klein-Gordon oscillators \cite{re24,FF2}. Therefore, it seems interesting to investigate the generalized Dirac oscillator associated with a neutral particle having a permanent magnetic dipole moment interacting with the external electromagnetic fields in $(2+1)$-dimensional cosmic string space-time. Our main goal is to consider the Dirac oscillator with a Coulomb-type potential function.

 The outline of this paper is:
In Sect. \eqref{sec2}, we investigate the influences of the A-C and curvature effects on the Dirac oscillator containing coulomb-type potential function and obtain the corresponding relativistic radial wave equation. In Sect. \eqref{sec3}, after replacing the Coulomb-type potential function in the resulting wave equation, we try to solve it and find the relativistic exact analytical solutions by applying the Nikiforov-Uvarov (NU) method. In Sect. \eqref{Conc}, we present the conclusions
\section{The generalized Dirac oscillator with the Aharono-Casher effect in cosmic string space-time \label{sec2}}
The influence of topological defects on quantum effect has been widely studied in several works \cite{re28,re29}, and the cosmic string is known as a simple example of a defect that contains curvature. Accordingly, we are motivated in investigating the influence of the A-C effect on the interaction of a Coulomb-type nonminimal coupling with the spinor field related to a neutral particle interacting with the external electric field in $(1+2)$-dimensional curved space-time. Accordingly, we begin by presenting the relevant line element as follows \cite{re30,re32} (in natural units $\hbar=c=G=1$)
\begin{equation}\label{LECosmicSt}
ds^{2}= d t^{2}-d r^{2}-\alpha^{2} r^{2} d \phi^{2},
\end{equation}
with $t\in(-\infty,\infty)$,  $r\in[0,\infty)$ and $\phi\in[0,2\pi]$. The parameter $\alpha$ is associated with the linear mass density, $\tilde{m}$, of the string through the relation $\alpha=1-4\tilde{m}$ and it runs in the range $(0,1]$ as well as it related to a deficit angle via $\gamma=2\pi(1-\alpha)$ \cite{re10}. If the $\alpha$ parameter run in the range $(1,\infty)$, it corresponds to an anti-conial space-time with no-positive curvature.
Nevertheless, the cosmic string characterized by equation 1 has a conical singularity that can be demonstrated by the curvature tensor $R_{r, \phi}^{r, \phi}=\frac{1-\alpha}{4 \alpha} \delta_{2}(\mathbf{r})$, with two-dimensional Dirac delta $\delta_{2}(\mathbf{r})$. Based on the line element in Eq. \eqref{LECosmicSt}, the metric tensor $g_{\mu\nu}$ and inverse metric tensor $g^{\mu\nu}$ indicate

\begin{equation}
	g_{\mu\nu}=\left(\begin{array}{cccc}
		1 & 0 & 0 \\
		0 & -1 & 0  \\
		0 & 0 & -\alpha^{2}r^{2}
	\end{array}\right),  g^{\mu\nu}=\left(\begin{array}{cccc}
		1 & 0 & 0 \\
		0 & -1 & 0 \\
		0 & 0 & -\frac{1}{\alpha^{2}r^{2}}
	\end{array}\right).
\end{equation}

The covariant Dirac equation related to dynamics of a neutral spin-half particle, with the permanent magnetic dipole moment, $\tilde{\mu}$, reads \cite{re33}
\begin{equation}\label{GDirac1}
	\left[i \gamma^{\mu}(x)\left(\partial_{\mu}+\Gamma_{\mu}(x)\right)+\frac{\tilde{\mu}}{2} \sigma^{\mu \nu}(x) F_{\mu \nu}-M\right] \Psi(t, \mathbf{r})=0,
\end{equation}
where $\Gamma_{\mu}(x)=\frac{1}{8} \omega_{\mu a b}(x) [\gamma^{a},\gamma^{b}]$ is named as spinor affine connection, $M$ denote mass of the particle, the generalized Dirac matrices, $\gamma^{\mu}=e^{\mu}_{\,\,\,a}\gamma^{a}$, are associated with the ordinary Dirac matrices, $\gamma^{0}=\sigma^{3}$, $\gamma^{1}=i\sigma^{1}$, $\gamma^{2}=i\sigma^{2}$, in $(1+2)$-dimensional flat space-time. They satisfy the anticommutation relation  $\left\{\gamma^{a}, \gamma^{b}\right\}=2 \eta^{ab}$. In addition, the electromagnetic tensor is denoted by $F_{\mu \nu}$ in such a way that one can demonstrate its components as
$F_{0i}=-F_{i0}=E_{i}$ and $F_{ij}=\epsilon_{ijk}B^{k}$. The Dirac oscillator with the oscillator frequency $\omega$ can be given by non-minimal substitution\ $\mathrm{p}_{\mu} \rightarrow\mathrm{p}_{\mu}+i M \omega \beta \mathrm{x}_{\mu}$ with $\beta=\gamma^{0}$\cite{re1,re2}. Besides, the generalized Dirac oscillator can be written according to the following modified momentum operator \cite{re20}
\begin{equation}\mathrm{p}_{\mu} \longrightarrow \mathrm{p}_{\mu}+i M \omega \beta \mathrm{X}_{\mu}\delta^{r}_{\mu}, \qquad \mathrm{X}_{\mu} \rightarrow \mathrm{X}_{r}\equiv f(r).
\end{equation}
In this case, by considering the non-minimal coupling, the generalized Dirac oscillator with arbitrary potential function in this system reads

\begin{equation}\label{GDirac2}
\begin{aligned}
&\left[i \gamma^{\mu}(x)\left(\partial_{\mu}+\Gamma_{\mu}(x)+M \omega f(r) \gamma^{0} \delta_{\mu}^{r}\right) \right]\Psi(t,{\bf r})\\
&+\left[\frac{\tilde{\mu}}{2} \sigma^{\mu \nu}(x) F_{\mu \nu}-M\right]\Psi(t,{\bf r})=0.
\end{aligned}
\end{equation}
We choose the basis tetrad in the cosmic string background as follow:
\begin{equation}
	e_{\,\,\,a}^{\mu}=\left(\begin{array}{cccc}
		1 & 0 & 0 \\
		0 & \cos \phi & \sin \phi  \\
		0 & -\frac{\sin \phi}{\alpha r} & \frac{\cos \phi}{\alpha r}
	\end{array}\right), e_{\,\,\,\mu}^{a}=\left(\begin{array}{cccc}
		1 & 0 & 0 \\
		0 & \cos \phi & -\alpha r \sin \phi \\
		0 & \sin \phi & \alpha r \cos \phi
	\end{array}\right).
\end{equation}
Based on properties of the basis tetrad, the $\gamma^{\mu}(x)$ matrices can be expressed as
\begin{eqnarray}\label{Ggamma}
	&&\gamma^{t}=\gamma^{0},\quad  \gamma^{r}(x)=\tilde{\gamma}^{r}(x)=\cos \phi \gamma^{1}+\sin \phi \gamma^{2}, \nonumber\\ &&\gamma^{\phi}(x)=\frac{\tilde{\gamma}^{\phi}}{\alpha r}=\frac{(-\sin \phi \gamma^{1}+\cos \phi \gamma^{2})}{\alpha r},\nonumber\\  && \gamma^{\phi}(x) \Gamma_{\phi}(x)=\frac{(1-\alpha)}{2 \alpha r} \gamma^{\,r}.
\end{eqnarray}
On the other hand, the electric field consists of two distinct parts, one of which is denoted by $\textbf{E}_{1}$, which comes from a uniformly infinite charge filament located along the $z$-axis perpendicular to the polar plane, and the other indicated by $\textbf{E}_{2}$ originated of a uniformly charged non-conducting cylinder whose length and radius are $L$ and $R$, respectively \cite{re34,re35}.
It is worth mentioning that
the following configuration of the electric field can lead to the appearance of the A-C effect in this framework,
\begin{subequations}
\begin{align}
&\textbf{E}_{1}=\frac{2 \lambda_{1}}{r} \widehat{e}_{r},\quad \nabla \cdot \textbf{E}_{1}=2 \lambda_{1} \delta(r), \quad\left(r=\sqrt{x^{2}+y^{2}}\right),\label{E1}\\
&\textbf{E}_{2}=\frac{\lambda_{2} r}{2} \widehat{e}_{r}, \quad \nabla \cdot \textbf{E}_{2}=\lambda_{2},  \frac{\partial \textbf{E}_{2}}{\partial t}=0,   \nabla \times \textbf{E}_{2}=0,\label{E2}
\end{align}
\end{subequations}
with $\lambda_{1}=\lambda_{0} / 4 \pi \epsilon_{0}, \quad  \lambda_{2}=\chi / \epsilon_{0}$,  the electric charge linear density $\lambda_{0}$ satisfies $\lambda_{0}>0$ , $r>0$ is the radial coordinate, and the electric charge volumetric density of cylinder is $\chi=\left(Q / \pi R^{2} L\right)>0$. In the polar coordinates system $(t,r,\phi)$, by considering the non-minimal coupling of the generalized Dirac oscillator, obtaining the term corresponding to the interaction of a spin-half particle having a permanent magnetic dipole moment with the electric field arising from $\frac{\tilde{\mu}}{2} \sigma^{\mu \nu} F_{\mu \nu}=-\tilde{\mu} \tilde{\gamma}^{\phi}E_{r}$, and substituting Eq. \eqref{Ggamma} in Eq. \eqref{GDirac2}, we get

\begin{equation}\label{GDirac3}
\begin{aligned}
&\left[i\gamma^{t}\partial_{t}+i\tilde{\gamma}^{r}\partial_{r}+\frac{i}{\alpha r} \tilde{\gamma}^{\phi}\partial_{\phi}+\frac{1-\alpha}{2\alpha r}\tilde{\gamma}^{r}
+M\omega f(r)\tilde{\gamma}^{\phi} \right]\Psi(t,{\bf r})\\
&-\left(\tilde{\mu} \tilde{\gamma}^{\phi}E_{r}+M\right)\Psi(t,{\bf r})=0.
\end{aligned}
\end{equation}
Equation \eqref{GDirac3} is obtained in the absence of the magnetic field.

By applying the similarity transformation matrix $S(\phi)$ on the matrices $\tilde{\gamma}^{r}$ and $\tilde{\gamma}^{\phi}$, they can be reduced to the constant Dirac matrices as follows:
\begin{equation}\label{similmatrix}
S(\phi)\tilde{\gamma}^{r}S^{-1}(\phi)=\gamma^{1}\equiv i\sigma^{1}, S(\phi)\tilde{\gamma}^{\phi}S^{-1}(\phi)=\gamma^{2}\equiv i\sigma^{2}.
\end{equation}
With respect to Eqs. \eqref{Ggamma}, \eqref{E1}, \eqref{E2} and \eqref{similmatrix}, we can rewrite Eq. \eqref{GDirac3} in the following form
\begin{equation}\label{GDriac4}
\begin{aligned}
&\left[\sigma^{2}\left(-\frac{\partial_{\phi}}{\alpha r}+iM\left(\omega f(r)-\frac{\omega_{AC}}{2}\right)-\frac{i \Phi_{AC}}{2\pi \alpha r}\right) \right]\Psi(t,{\bf r})\\
&+\left[i \sigma^{3} \partial_{t}-\sigma^{1}\left(\partial_{r}+\frac{1-\alpha}{2\alpha r}\right)-M I\right)]\Psi(t,{\bf r})=0,
\end{aligned}
\end{equation}
where $I$ is a $2 \times 2$ identity matrix and
the parameters $\omega_{AC}\equiv \tilde{\mu}\lambda_{2}/M$ denotes the A-C frequency related to the spin-half particle. In this way, $\Phi_{AC}=4 \pi \alpha \tilde{\mu} \lambda_{1}$ indicates the A-C phase. Note that we deal with the modified electric field $\mathbf{E} \rightarrow \mathbf{E}/\alpha$
impressed by the cosmic string space-time. In this case, we assume the Dirac spinor reads \cite{re37,re38}
\begin{equation}\label{spinorDirac}
	\Psi(t, r, \theta)=\frac{e^{i\left(m_{l} \phi-E t\right)}}{\sqrt{2 \pi}}\begin{pmatrix}
		\psi_{+}(r) \\
		i \psi_{-}(r)
	\end{pmatrix}, \left(m_{l}=\pm 1 / 2, \ldots\right),
\end{equation}
Then, we can obtain the two coupled Dirac equations by substituting Eq. \eqref{spinorDirac} into Eq. \eqref{GDriac4}
\begin{equation} \label{hh1}
\begin{aligned}
&i\left[\partial_{r}+\frac{1-\alpha}{2 \alpha r}+\frac{m_{l}}{\alpha r}-M\left(\omega f(r)-\frac{\omega_{A C}}{2}\right)+\frac{\Phi_{A C}}{2\pi \alpha r}\right]\psi_{-}(r)\\
&=\left(E-M\right)\psi_{+}(r),
\end{aligned}
\end{equation}

\begin{equation}\label{hh2}
\begin{aligned}
&\left[-\partial_{r}-\frac{1-\alpha}{2 \alpha r}+\frac{m_{l}}{\alpha r}-M\left(\omega f(r)-\frac{\omega_{A C}}{2}\right)+\frac{\Phi_{A C}}{2\pi \alpha r}\right]\psi_{+}(r)\\
&=i\left(E-M\right)\psi_{-}(r).
\end{aligned}
\end{equation}
Afterward, to combine the two coupled Dirac equations to get the following second-order differential equation, we need to remove $\psi_{-}(r)$ from Eqs. \eqref{hh1} and \eqref{hh2} and arrive at
\begin{equation} \label{GDriac6}
\begin{split}
&\frac{d^2\psi_{+}(r)}{dr^2}+\frac{1-\alpha}{\alpha r}\frac{d\psi_{+}(r)}{dr}+\left[E^2-\xi+M \omega \frac{df(r)}{dr}\right.\\
&\left.+2M\left(\frac{m_{l}}{\alpha r}+\frac{\Phi_{AC}}{2\pi \alpha r}\right) \left(\omega f(r)-\frac{\omega_{AC}}{2}\right)+\left(\left(\frac{1}{2\alpha}-1\right)^2\right.\right.\\
&\left.\left.+\left(\frac{m_{l}}{\alpha }+\frac{\Phi_{AC}}{2\pi \alpha }\right)-\frac{1}{4}-\left(\frac{m_{l}}{\alpha }+\frac{\Phi_{AC}}{2\pi \alpha }\right)^2\right)\frac{1}{r^2}
\right]\psi_{+}(r)=0
\end{split}
\end{equation}
with $\xi=M^2+M^2\left(\omega f(r)-\frac{\omega_{AC}}{2}\right)^2$.
In this case, the generalized Dirac oscillator with the external electric field in the cosmic string background breaks the degeneracy \cite{re19} if the potential function $f(r)=r$ is embedded in the Dirac oscillator.
Our next goal is mainly to study the energy eigenvalues
related to the Dirac oscillator with the Coulomb-type potential function under the background of cosmic string space-time.
\section{Exact solutions under the Coulomb-type potential function f(r)} \label{sec3}
In recent years, the Coulomb potential has received much attention for its importance in many subjects of physics \cite{re25,QuesneJPAMG2004,HassanabadiGRG2018,re21,ZareIJMPA2020,MontignyEPJP2022,re44,re45}. From now on, by selecting the Coulomb form for the potential function, that is,
\begin{equation}\label{PotFunc}
f(r) = \frac{\mathrm{N}_{1}}{r},
\end{equation}
where the constant potential parameter $\mathrm{N}_{1}$ is related to the Coulomb strength, we are able to solve the second-order differential equation, given by Eq. \eqref{GDriac6}, by employing the NU method. Now, let us place the potential function given by Eq. \eqref{PotFunc} in the Eq. \eqref{GDriac6} and get to the following expression
\begin{equation} \label{GDriac7}
\begin{split}
&\frac{d^2\psi_{+}(r)}{dr^2}+\frac{1-\alpha}{\alpha r}\frac{d\psi_{+}(r)}{dr}
+\frac{1}{r^2}\left[-\tau_{1}\,r^2+\tau_{2}\,r-\tau_{3}\right]
\psi_{+}=0.\\
&\tau_{1}=-E^2+M^2\left(1+\frac{\omega^{2}_{AC}}{4}\right),\\
&\tau_{2}=M^2 \omega \omega_{AC} N_{1}-\left(\frac{m_{l}}{\alpha }+\frac{\Phi_{AC}}{2\pi \alpha }\right)M\omega_{AC},\\
&\tau_{3}=\frac{1}{4}-\left(\frac{1}{2\alpha}-1\right)^2-\left(\frac{m_{l}}{\alpha }+\frac{\Phi_{AC}}{2\pi \alpha }\right)+\left(\frac{m_{l}}{\alpha }+\frac{\Phi_{AC}}{2\pi \alpha }\right)^2\\
&\quad+M\omega N_{1}\left(1+M\omega N_{1}-2\left(\frac{m_{l}}{\alpha}+\frac{\Phi_{AC}}{2\alpha \pi}\right)\right).
\end{split}
\end{equation}
Obviously, Eq. \eqref{GDriac7} is a particular case of the NU equation \cite{HassanabadiGRG2018,re40,re41, MontignyEPJP2022,TezcanIJTP2009,MontignyGRG2018,MontignyEPJP2021}. It is well-known that the NU method is an efficient model for finding eigenfunctions and eigenvalues of the Schr\"odinger-like equation \cite{MontignyEPJP2022}. Therefore, the second-order differential equation that is suitable for implementing the NU method can be presented as

\begin{equation}\label{NUEq}
\begin{aligned}
&\left[\frac{d^{2}}{d r^{2}}+\frac{1}{\left[r\left(1-\beta_{3} r\right)\right]^{2}}\left(-\zeta_{1} r^{2}+\zeta_{2} r-\zeta_{3}\right)\right]\psi(r)\\
&+\frac{\beta_{1}-\beta_{2} r}{r\left(1-\beta_{3} r\right)} \frac{d}{d r}\psi(r)=0,
\end{aligned}
\end{equation}
whose eigenfunctions can be written as
\begin{equation}
\psi(r)=r^{\beta_{12}}\left(1-\beta_{3}r\right)^{-\beta_{12}-\frac{\beta_{13}}{\beta_{3}}}P_{n}^{\left(\beta_{10}-1,\frac{\beta_{11}}{\beta_{3}}-\beta_{10}-1\right)}\left(1-2\beta_{3}r\right),
\end{equation}
also its eigenvalues can be presented by the following equation
\begin{equation}
\begin{split}
&n \beta_{2}-(2 n+1) \beta_{5}+(2 n+1)\left(\sqrt{\beta_{9}}-\beta_{3} \sqrt{\beta_{8}}\right) \nonumber\\
&+\beta_{7}+2 \beta_{3} \beta_{8}+2 \sqrt{\beta_{8} \beta_{9}}+n(n-1) \beta_{3}=0.
\end{split}
\end{equation}
Note that the parameters $\beta_{4}\dots, \beta_{13}$ can be found from the six parameters $\beta_{1}, \beta_{2}, \beta_{3}$, $\zeta_{1}, \zeta_{2}$ and $\zeta_{3}$ in Eq. \eqref{NUEq} according to the following relation (see Refs. \cite{MontignyEPJP2022,re40,TezcanIJTP2009,MontignyGRG2018,MontignyEPJP2021} for more details.)
\begin{equation}
\begin{split}
&\beta_{4}=\frac{1}{2}\left(1-\beta_{1}\right), \beta_{5}=\frac{1}{2}\left(\beta_{2}-2 \beta_{3}\right),  
\beta_{6}=\beta_{5}^{2}+\zeta_{1}, \\
&\beta_{7}=2 \beta_{4} \beta_{5}-\zeta_{2},  \beta_{8}=\beta_{4}^{2}+\zeta_{3}, 
\beta_{9}=\beta_{3}\beta_{7}+\beta_{3}^{2}\beta_{8}+\beta_{6},\\
&\beta_{10}=\beta_{1}+2\beta_{4}+2\sqrt{\beta_{8}}, \\
&\beta_{11}=\beta_{2}-2\beta_{5}+2\left(\sqrt{\beta_{9}}+\beta_{3}\sqrt{\beta_{8}}\right),\\
&\beta_{12}=\beta_{4}+\sqrt{\beta_{8}}, \beta_{13}=\beta_{5}-\sqrt{\beta_{9}}-\beta_{3} \sqrt{\beta_{8}}.
\end{split}
\end{equation}
If we focus on Eq. \eqref{GDriac7}, we see that it is a particular case of the differential equation compatible with the NU method, where $\beta_{3}=0$. So, it makes sense that
\begin{equation}\label{limitedpsi}
\begin{split}
\lim\limits_{\beta_{3} \rightarrow 0} \psi(r)=r^{\beta_{12}} e^{\left(\beta_{13} r\right)} L_{n}^{\left(\beta_{10}-1\right)}\left(\beta_{11} r\right).
\end{split}
\end{equation}
Given Eq. \eqref{limitedpsi}, it can be seen that the $\psi(r)$ function is related to the generalized Laguerre polynomial, $L_{n}^{\beta} (x)$. With regard to these descriptions, the eigenfunctions related to the wave equation given by Eq. \eqref{GDriac7} can be written as follows
\begin{equation}\label{WF1}
\psi_{+}(r)=r^{1-\frac{1}{2\alpha}+\sqrt{\tau_{3}+\left(\frac{1}{2\alpha}-1\right)^{2}}}\,e^{-\sqrt{\tau_{1}}}\,\,L_{n}^{2\sqrt{\tau_{3}+\left(\frac{1}{2\alpha}-1\right)^{2}}}\left(2\sqrt{\tau_{1}}\, r\right),
\end{equation}
and the corresponding energy eigenvalues can be determined as
\begin{equation}\label{energy1}
\begin{split}
E_{n\,m_{l}}=\pm \sqrt{M^{2}\left(1+\frac{\omega^{2}_{AC}}{4}\right)-\frac{\tau^{2}}{\left[1+2n+2\sqrt{\tau_{3}+\left(\frac{1}{2\alpha}-1\right)^2}\right]^{2}}}\,.
\end{split}
\end{equation}
\begin{figure}
\includegraphics[width=0.46\textwidth]{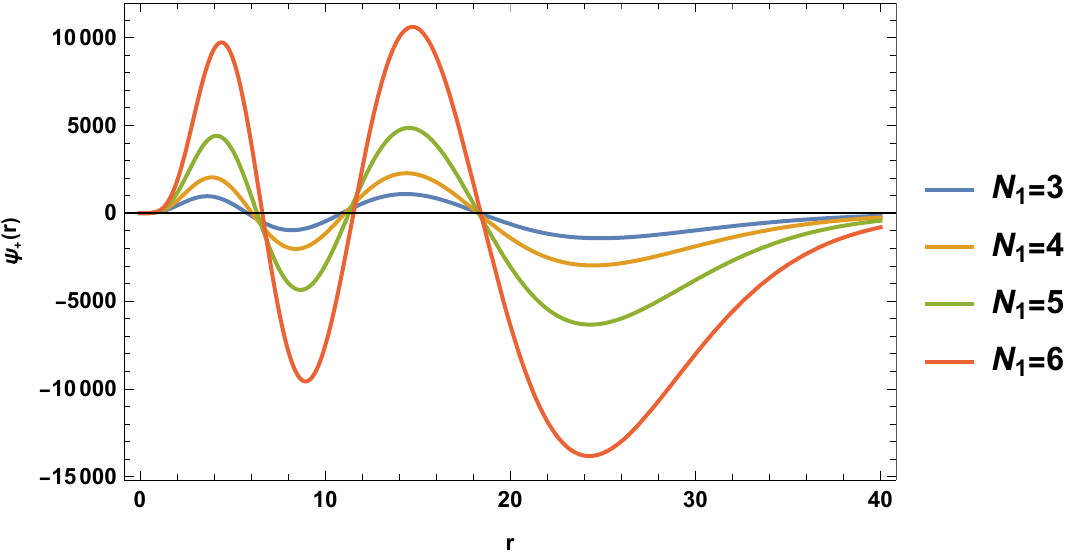}
\caption{The eigenfunctions in Eq. \eqref{WF1} as a function of $r$ in terms of four different values of $N_{1}$ and other parameters $n=3$, $\omega=M=m_{l}=\lambda_{2}=1$, $\lambda_{1}=-1$, $\tilde{\mu}=2$ and $\alpha=0.8$.}
\label{FIG1}
\end{figure}
\begin{figure*}
\resizebox{\linewidth}{!}{\includegraphics{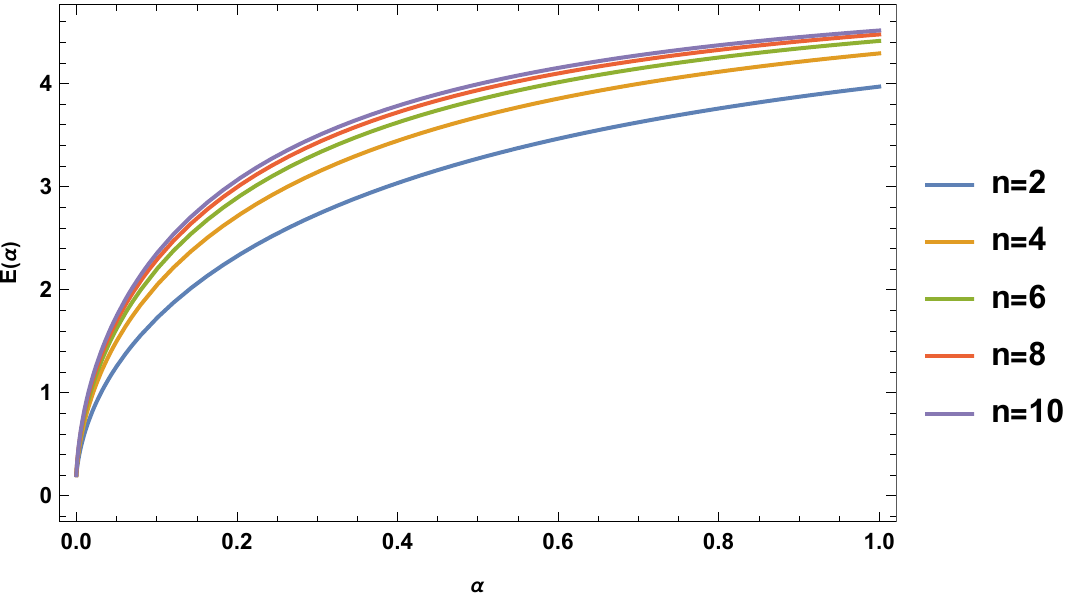},\includegraphics{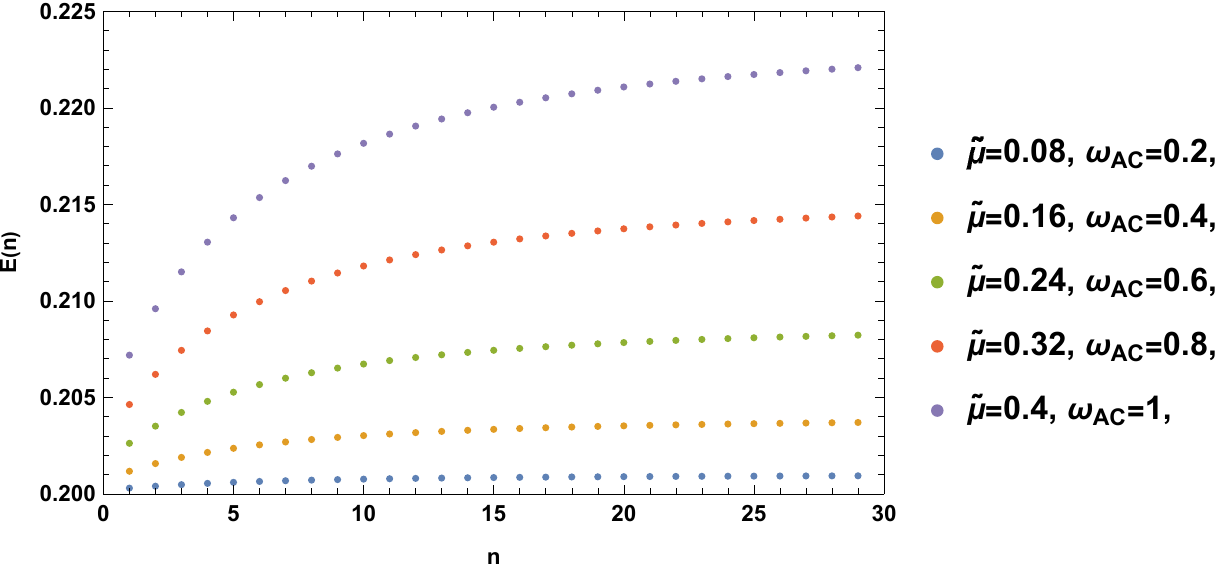}}
\caption{ Left panel: The energy eigenvalues in Eq. \eqref{energy1} as a function of $\alpha$ in terms of five different values of the quantum number $n$ and other parameters $\omega=1$, $M=0.2$, $m_{l}=2$, $N_{1}=-3$, $\lambda_{1}= -0.01$, $\lambda_{2}=5$ and $\tilde{\mu}=2$. Right
panel:The energy eigenvalues in Eq. \eqref{energy1} as a function of $n$ in terms of five different values of the A-C frequency parameter $\omega_{AC}$ and the permanent magnetic dipole moment parameter $\tilde{\mu}$, and other parameters $\omega=1$, $M=0.2$, $m_{l}=2$, $N_{1}=-3$, $\lambda_{1}= -0.01$, $\lambda_{2}=0.5$ and $\alpha=0.2$.}\label{fig:GreyBodyfromA0}
\end{figure*}
\begin{figure*}
\resizebox{\linewidth}{!}{\includegraphics{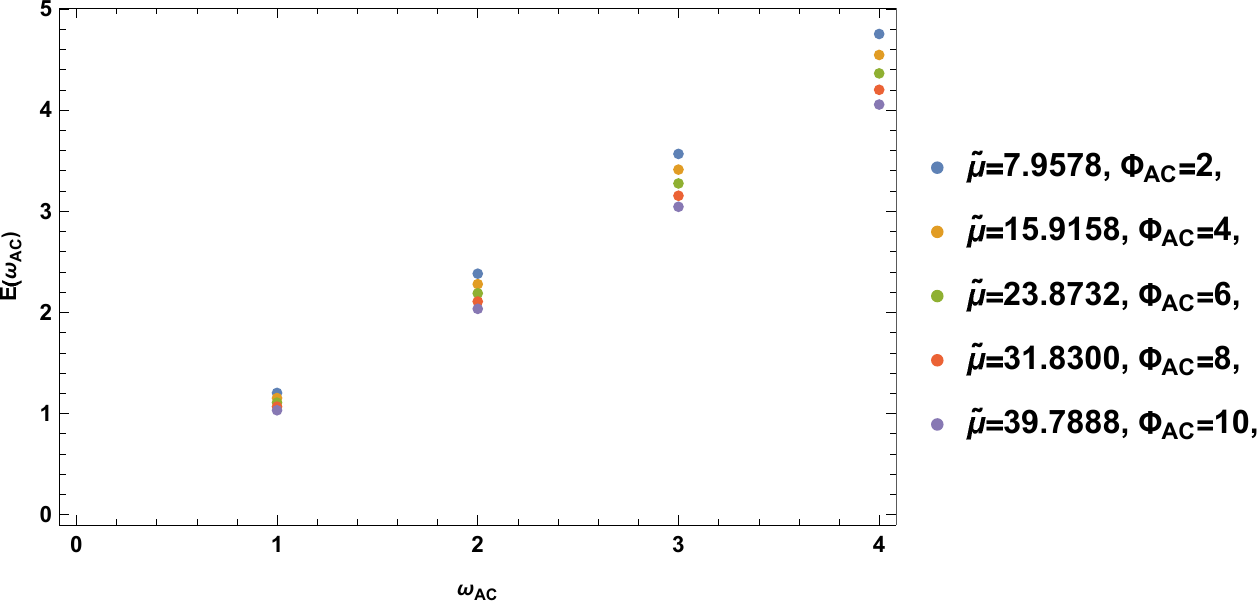},\includegraphics{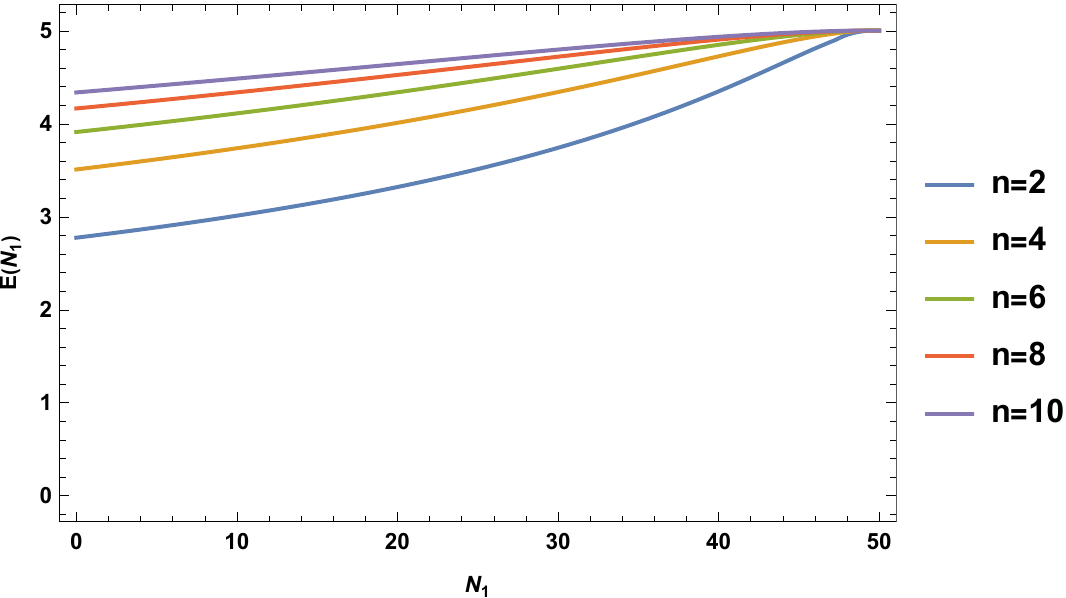}}
\caption{Left panel: The energy eigenvalues in Eq. \eqref{energy1} as a function of he A-C frequency parameter $\omega_{AC}$ in terms of five different values of the A-C geometric phase $\Phi_{AC}$ and the permanent magnetic dipole moment parameter $\tilde{\mu}$, and other parameters $\omega=1$, $M=0.2$, $m_{l}=2$, $N_{1}=-3$, $\lambda_{1}= 0.1$, $\lambda_{2}=0.5$, $\alpha=0.2$ and $n=3$. Right
panel: The energy eigenvalues in Eq. \eqref{energy1} as a function of the Coulomb strength $N_{1}$ in terms of five different values of the quantum number $n$ and other parameters $\omega=1$, $M=0.2$, $m_{l}=2$, $\lambda_{1}= -0.01$, $\lambda_{2}=5$ and $\alpha=0.2$.}\label{fig:GreyBodyfromA0}
\end{figure*}
In Eq. \eqref{energy1}, one can see the influence of the Aharonov-Casher effect on the relativistic energy spectrum of the generalized Dirac oscillator containing the coulomb-type potential function associated with a neutral fermion with the permanent magnetic dipole interacting with the external electric field in $(1+2)$-dimensional cosmic string spacetime. Therefore it is clear that the relevant energy spectrum depend on the the A-C frequency parameter $\omega_{AC}$, the A-C geometric phase $\Phi_{AC}$, the Coulomb strength $N_{1}$, the defect parameter $\alpha$ related to the deficit angle, the oscillator frequency $\omega$, the mass of the fermion $M$, and quantum numbers $n$ and $m_{l}$.

Similarly, it can be seen that, under the influences of the A-C and curvature effects on such a Dirac oscillator in the present external electric field, the corresponding eigenfunctions depend on defined parameters $\omega_{AC}$, $\Phi_{AC}$, $\alpha$, $N_{1}$, $\omega$, $n$ and  $m_{l}$.

Now, to observe the influence of the Coulomb strength $N_{1}$ on the radial wave function given by  Eq. \eqref{WF1}, we draw $\psi_{+}(r)$ in terms of the coordinates $r$, according to four different values of the Coulomb strength $N_{1}$ in FIG.1.

It is work worth emphasizing that the energy eigenvalues of the Dirac oscillator is obtained for the ordinary potential function, that is, $f(r)=r$ in Ref. \cite{re19}.
To visually observe the influence of various parameters on the energy spectrum in the system, we consider the positive energies states. Therefore, in FIG.2(Left panel), we show that the relevant energy eigenvalues considered as the function of $\alpha$ in terms of five different values of $n$ can increase with a decreasing slope. Similarly, as the values of the A-C frequency $\omega_{AC}$ increases, the energy spectrum also gets smaller in FIG.2(Right panel).

Furthermore, we can intuitively understand the influence of the external electric fields on the energy spectrum of the generalized Dirac oscillator containing the Coulomb-type potential function in FIG.3(Left panel). Thus, we indicate that the corresponding energy eigenvalues, considered as the function of $\omega_{AC}$ in terms of five different values of $\tilde{\mu}$ and $\Phi_{AC}$, decrease as the values of the permanent magnetic dipole moment parameter $\tilde{\mu}$
	and the A-C geometric phase $\Phi_{AC}$ increase.
	It should be noted that the A-C phase $\Phi_{AC}$ implies a
	periodicity on the relevant energy eigenvalues so that
	$E_{n m_{l}}\left(\Phi_{A C} \pm 2 \pi\right)=E_{n m_{l}+1}\left(\Phi_{A C}\right)$ \cite{re16}.
	Besides, one can see that the relevant relativistic energy eigenvalues considered as a function of $N_{1}$ increase as the quantum number $n$ increases in FIG.3(Right panel).
	It is clear that the energy eigenvalues of the relevant generalized Dirac oscillator are modified by the Coulomb strength parameter $N_{1}$ related to the Coulomb-type potential function.
\section{Conclusions}\label{Conc}
In the framework of quantum field theory in curved space-time, we try to investigate the influence of the curvature and the A-C effects on the interaction of a Coulomb-type nonminimal coupling with the spinor field related to a relativistic neutral fermion interacting with the external electric field in $(1+2)$-dimensional cosmic string space-time. Meanwhile, the associated particle possesses a permanent magnetic dipole moment in the direction of the z-axis, which interacts with the external electric field constructed by two distinct radial field configurations; the first comes from a uniformly infinite charge filament presented by a Coulomb-type electric field, and the other originated of a uniformly charged non-conducting cylinder denoted by a linear-type electric field. Then, in the middle of solving the generalized Dirac oscillator containing the Coulomb-type potential function to find the exact analytical solutions related to the particle, we see that the Coulomb-type electric field leads to appears the A-C geometric quantum phase, and the linear-type electric field provides the A-C frequency in the associated wave equation.

In this way, by applying the NU method, we solve the corresponding generalized Dirac oscillator and obtain the eigenfunction and relevant energy eigenvalues.
Then, we see that the resulting solutions are shifted because of the Coulomb-potential function, A-C, and curvature effects, as can be seen in dependence of the A-C frequency parameter $\omega_{AC}$, the A-C geometric phase $\Phi_{AC}$, the Coulomb strength $N_{1}$, the defect parameter $\alpha$ related to the deficit angle and the oscillator frequency $\omega$. By the way, we can see that the degeneracy of the relevant relativistic energy eigenvalues is broken by depending on the Coulomb strength parameter under the influence of the curvature effect and the Aharonov-Casher effect. In addition, we plot some figures of the resulting solutions for further examination.
\vspace{5mm}
\begin{acknowledgments}
This work is supported by the National Natural Science Foundation of China (Grant Nos. 11465006 and 11565009) and the Major Research Project of innovative Group of Guizhou province (2018-013).
\end{acknowledgments}

\end{document}